\documentclass{article}
\usepackage{amsfonts}
\usepackage{amsmath}

\newtheorem{theorem}{Theorem}

\newtheorem{corollary}[theorem]{Corollary}

\newtheorem{definition}[theorem]{Definition}

\newtheorem{lemma}[theorem]{Lemma}

\newtheorem{problem}[theorem]{Problem}
\newtheorem{proposition}[theorem]{Proposition}

\newenvironment{proof}[1][Proof]{\noindent\textbf{#1.} }{\ \rule{0.5em}{0.5em}}
\input{tcilatex}
\begin{document}

\title{Upper and lower bounds on dynamic risk indifference prices in
incomplete markets}
\author{Xavier De Scheemaekere\thanks{%
F.R.S.-F.N.R.S. Research Fellow; Universit\'{e} Libre de Bruxelles (ULB),
Solvay Brussels School of Economics and Management, Centre Emile Bernheim;
Postal adress: Av. F.D. Roosevelt, 50, CP 145/1, 1050 Brussels, Belgium ;
Phone: +32.2.650.39.59; E-mail: xdeschee@ulb.ac.be}}
\maketitle

\begin{center}
\bigskip \textbf{Abstract}
\end{center}

In the context of an incomplete market with a Brownian filtration and a
fixed finite time horizon $T$, this paper proves that for general dynamic
convex risk measures, the buyer's ($p_{t}^{buyer}$) and seller's ($%
p_{t}^{seller}$) risk indifference prices of a bounded contingent claim
satisfy 
\begin{equation*}
p_{t}^{low}\leq p_{t}^{buyer}\leq p_{t}^{seller}\leq p_{t}^{up},\text{\ }%
\forall t\in \lbrack 0,T],
\end{equation*}%
where $p_{t}^{low}$ and $p_{t}^{up}$ are the dynamic lower and upper hedging
prices, respectively$.$

\bigskip


\bigskip


\bigskip


\section{Introduction}

In incomplete markets, arbitrage-free pricing of contingent claims is not
unique. The no-arbitrage assumption provides infinitely many equivalent
martingale measures and yields an interval of arbitrage-free prices, instead
of a unique price (Harrison and Pliska \cite{HP}, Delbaen and Schachermayer 
\cite{delbaen and sacher}). The reason is that perfect replication is
impossible and risk cannot be fully eliminated.

Still, upper and lower hedging prices (El Karoui and Quenez \cite{El Karoui
and Quenez}, Kramkov \cite{Kramkov}) can be charged in order to eliminate
all risks. The upper hedging price represents the minimal initial payment
needed for the hedging portfolio to attain a terminal wealth that is no less
than the derivative payoff.\ This price, however, is excessively high, as it
often reduces to the trivial upper bound of the no-arbitrage interval
(Eberlein and Jacod \cite{eberlein jacod}, Bellamy and Jeanblanc \cite%
{Bellamy and jeanblanc}). In order to get more information on the asset
value, one possibility is to introduce an optimality criterion that puts
more restrictions on the bounds of the price interval.

A few examples include picking martingale measures according to optimal
criteria (Bellini and Frittelli \cite{bellini and fritelli}, F\"{o}llmer and
Schweizer \cite{follmer and schweizer}, Gerber and Shiu \cite{gerber and
shiu}, Goll and R\"{u}schendorf \cite{goll and ruschendorf}), invoking
(exponential) utility indifference arguments (Ankirchner et al. \cite%
{Ankirchner Imkeller Poppier}, Becherer \cite{Becherer}, El Karoui and Rouge 
\cite{EK and Rouge}, Henderson and Hobson \cite{18}, Musiela and
Zariphopolou \cite{24}), using dynamic risk measures for the optimal design
of derivatives (Barrieu and El Karoui \cite{Barr et El K I},\cite{Bar et El
K II},\cite{4}), pricing by stress measures (Carr et al. \cite{carr et al}),
or good-deal asset price bounds (Cochrane and Sa\`{a}-Requejo \cite{cochrane}%
), among many others $-$ more details can be found in Xu \cite{29}, or Horst
and M\"{u}ller \cite{Horst and Muller}, and the references there in.\ 

In this work, the optimality criterion comes from the risk indifference
principle, recently proposed for pricing in incomplete markets (Kl\"{o}ppel
and Schweizer \cite{K&S II}, \O ksendal and Sulem \cite{25}, Xu \cite{29}).
The (seller's) dynamic risk indifference price is the initial payment that
makes the risk involved for the seller of a contract equal, at any time, to
the risk involved if the contract is not sold, with no initial payment.
Hence, the resulting \emph{price} is such that the agent is \emph{indifferent%
} between his \emph{risk} if a transaction occurs and his \emph{risk} if no
transaction occurs.

The abstract risk indifference pricing setting has been studied in Xu \cite%
{29}, where it is shown to generalize utility-based derivative pricing
introduced by Hodges and Neuberger \cite{19} and valuation by stress
measures of Carr et al. \cite{carr et al}.\ 

In a static Markovian framework, and for a particular class of convex risk
measures, \O ksendal and Sulem \cite{25} study the risk indifference method
in a jump diffusion market, using PDE methods. In particular, they prove
that the buyer's ($p^{buyer}$) and seller's ($p^{seller}$) risk indifference
prices satisfy 
\begin{equation}
p^{low}\leq p^{buyer}\leq p^{seller}\leq p^{up},  \label{fund ineq}
\end{equation}%
where $p^{low}$ and $p^{up}$ are the lower and upper hedging prices,
respectively.

In the context of a Brownian filtration, this paper shows that (\ref{fund
ineq}) holds for \textit{general} \textit{dynamic risk indifference prices}.
The result is based on a recent work of Delbaen et al. \cite{Delbaen et al},
who provide a representation of the penalty term of general dynamic convex
risk measures, and it is obtained by applying backward stochastic
differential equation (BSDE) theory.

The paper is organised as follows.\ Section 2 presents the financial market
model and recalls some well-known results on BSDEs and their application in
finance. Section 3 introduces the super-replication technique, and the risk
indifference principle, for pricing in incomplete markets.\ Section 4
contains the main result of the paper, that is, the proof that general
dynamic risk indifference prices are bounded from below and above by lower
and upper hedging prices, respectively.\ As we shall see, BSDE theory
provides a unified mathematical perspective on complete markets,
super-replication and indifference prices, and the relationships between
them.

\section{Notation and preliminaries}

\subsection{Financial market model}

Let a given $T\in (0,\infty )$ be the fixed finite time horizon of a
financial market with two investment possibilities:

(i) A risk free security (e.g. a bond), with unit price price $B_{t}=1$ at
all times $t\in \lbrack 0,T]$.

(ii) $n$ risky securities (e.g. stocks), with prices evolving according to
the following equation:%
\begin{equation*}
\frac{dS_{t}}{S_{t}}=\mu _{t}dt+\sigma _{t}dW_{t},\qquad S_{0}>0,\qquad
S_{t}\in 
%TCIMACRO{\U{211d} }%
%BeginExpansion
\mathbb{R}
%EndExpansion
^{n\times 1},\qquad t\in \lbrack 0,T].
\end{equation*}%
In the above, $(W_{t})_{t\in \lbrack 0,T]}$ is a standard $d$-dimensional
Brownian motion defined on a probability space $(\Omega ,\tciFourier ,P)$, $%
\left\{ \tciFourier _{t}\right\} _{t\in \lbrack 0,T]}$ is the augmented
filtration generated by $(W_{t})_{t\in \lbrack 0,T]},$ $\mu _{t}\in 
%TCIMACRO{\U{211d} }%
%BeginExpansion
\mathbb{R}
%EndExpansion
^{n\times 1}$ is a $\tciFourier _{t}$-predictable vector-valued map and $%
\sigma _{t}\in 
%TCIMACRO{\U{211d} }%
%BeginExpansion
\mathbb{R}
%EndExpansion
^{n\times d}$\ is a $\tciFourier _{t}$-predictable full rank matrix-valued
map. $W$ is described as a column vector of dimension $d\times 1,$ and $%
\sigma ^{i}dW$ has to be understood as a matrix product with dimension $%
1\times 1$ for each $1\leq i\leq n$.

Assume that the processes $\mu $ and $\sigma $\ satisfy%
\begin{equation*}
\dint\nolimits_{0}^{T}\left\vert \mu _{s}\right\vert +\left\Vert \sigma
_{s}\right\Vert ^{2}ds<\infty ,\text{ }P-a.s.
\end{equation*}

A portfolio in this market is modeled by the $1\times n$ row vector $\pi
_{t},$\ representing for the amount invested in the risky assets at time $t$%
. The dynamics of the corresponding wealth process $X(t)=X_{x}^{(\pi )}(t)$
is%
\begin{eqnarray*}
dX_{t} &=&\pi _{t}\frac{dS_{t}}{S_{t}}=\pi _{t}\left[ \mu _{t}dt+\sigma
_{t}dW_{t}\right] ;\qquad t\in \left[ 0,T\right] \\
X_{0} &=&x>0.
\end{eqnarray*}%
The portfolio $\pi _{t}$ is admissible if it is $\tciFourier _{t}$%
-predictable and satisfies

\begin{equation*}
\dint\nolimits_{0}^{T}\left( \left\vert \mu _{t}\right\vert \left\vert \pi
_{t}\right\vert S_{t}+\left\Vert \sigma _{t}\right\Vert ^{2}\left\vert \pi
_{t}\right\vert ^{2}\left\vert S_{t}\right\vert ^{2}\right) dt<\infty
\end{equation*}%
and

\begin{equation*}
X_{t}\geq 0\text{ \ for }t\in \lbrack 0,T],\text{ \ }P-a.s.
\end{equation*}%
Let $\Pi $ denote the set of all admissible portfolios.

Suppose that the $\tciFourier _{T}$-measurable random variable $\xi =\xi
_{T} $ represents a contingent claim with maturity $T>0$, depending on $%
S(\cdot ). $

When the market is complete, i.e. $n=d$, it is well known that the unique
dynamic arbitrage-free price of $\xi $ is given by the Black-Scholes price%
\begin{equation*}
p_{t}^{BS}:=E_{Q}[\xi |\tciFourier _{t}],
\end{equation*}%
where $Q=Q^{\theta }$ is the so-called unique equivalent martingale measure
defined as%
\begin{equation}
E_{P}\left[ \frac{dQ}{dP}|\tciFourier _{T}\right] :=\exp \left(
\dint\nolimits_{0}^{T}\theta _{s}dW_{s}-\frac{1}{2}\dint\nolimits_{0}^{T}%
\left\vert \theta _{s}\right\vert ^{2}ds\right) ,  \label{stoch exp}
\end{equation}%
and $(\theta _{t})_{t\in \lbrack 0,T]}$ is a suitable $%
%TCIMACRO{\U{211d} }%
%BeginExpansion
\mathbb{R}
%EndExpansion
^{1\times d}$-valued process, called the market price of risk, such that $%
\sigma _{t}\theta _{t}^{\top }+\mu _{t}=0$.

By Girsanov theorem, $\widetilde{W}_{t}:=W_{t}-\tint\nolimits_{0}^{T}\theta
_{s}ds$ is a $Q$-Brownian motion, and the (discounted) price process $S_{t}$
is a martingale with respect to $Q$ $-$ hence the name \textit{equivalent
martingale measure} (EMM for short).

In the sequel, we identify a probability measure $Q$ equivalent to $P$ with
its Radon-Nikodym density $\frac{dQ}{dP}$ and with the predictable process $%
(\theta _{t})_{t\in \lbrack 0,T]}$ induced by the stochastic exponential, as
in (\ref{stoch exp}).

\subsection{BSDEs and complete markets}

Consider the function 
\begin{equation*}
g:%
%TCIMACRO{\U{211d} }%
%BeginExpansion
\mathbb{R}
%EndExpansion
\times \Omega \times 
%TCIMACRO{\U{211d} }%
%BeginExpansion
\mathbb{R}
%EndExpansion
\times 
%TCIMACRO{\U{211d} }%
%BeginExpansion
\mathbb{R}
%EndExpansion
^{1\times d}\rightarrow 
%TCIMACRO{\U{211d} }%
%BeginExpansion
\mathbb{R}
%EndExpansion
:(t,\omega ,y,z)\rightarrow g(t,\omega ,y,z)
\end{equation*}%
satisfying the following assumptions (to simplify the notation, we often
write $g(t,y,z)$\ instead of $g(t,\omega ,y,z))$:

\begin{itemize}
\item[(i)] $g$ is Lipschitz in $(y,z)$, i.e. there exists a constant $u>0$
such that we have, $dt\times dP-a.s.$, for any $(y_{0},z_{0}),(y_{1},z_{1})%
\in 
%TCIMACRO{\U{211d} }%
%BeginExpansion
\mathbb{R}
%EndExpansion
\times 
%TCIMACRO{\U{211d} }%
%BeginExpansion
\mathbb{R}
%EndExpansion
^{1\times d},$%
\begin{equation*}
\left\vert g(t,y_{0},z_{0})-g(t,y_{1},z_{1})\right\vert \leq u\left(
\left\vert y_{0}-y_{1}\right\vert +\left\vert z_{0}-z_{1}\right\vert \right)
.
\end{equation*}

\item[(ii)] For all $(y,z)\in 
%TCIMACRO{\U{211d} }%
%BeginExpansion
\mathbb{R}
%EndExpansion
\times 
%TCIMACRO{\U{211d} }%
%BeginExpansion
\mathbb{R}
%EndExpansion
^{d},$ $g(\cdot ,y,z)$ is a predictable process such that for any finite $%
T>0,$ we have $E[\tint\nolimits_{0}^{T}(g(t,\omega ,y,z))^{2}dt]<+\infty $
for any $y\in 
%TCIMACRO{\U{211d} }%
%BeginExpansion
\mathbb{R}
%EndExpansion
$ and $z\in 
%TCIMACRO{\U{211d} }%
%BeginExpansion
\mathbb{R}
%EndExpansion
^{1\times d}.$
\end{itemize}

A BSDE is an equation of the type 
\begin{eqnarray*}
-dY_{t} &=&g(t,Y_{t},Z_{t})-Z_{t}dW_{t}, \\
Y_{T} &=&\xi ,
\end{eqnarray*}

where $\xi $ is a random variable in $L^{2}(\Omega ,\tciFourier _{T},P).$ In
1990, Pardoux and Peng \cite{Pardoux and Peng} showed that for a finite time
horizon $T>0,$ there exists a unique solution $(Y_{t},Z_{t})_{t\in \lbrack
0,T]}$ consisting of predictable stochastic processes (the former $%
%TCIMACRO{\U{211d} }%
%BeginExpansion
\mathbb{R}
%EndExpansion
$-valued, the latter $%
%TCIMACRO{\U{211d} }%
%BeginExpansion
\mathbb{R}
%EndExpansion
^{1\times d}$-valued) such that $E[\tint\nolimits_{0}^{T}Y_{t}^{2}dt]<+%
\infty $ and $E[\tint\nolimits_{0}^{T}\left\vert Z_{t}\right\vert
^{2}dt]<+\infty .$

El Karoui et al.\ \cite{9} showed that BSDE theory provides the mathematical
background of many problems in finance.\ In particular, in a complete
market, the dynamic arbitrage-free price $p_{t}^{BS}$\ of a contingent claim 
$\xi \in L^{2}(\Omega ,\tciFourier _{T},P)$\ is equal to the solution $%
Y_{t}^{BS}$ of the following BSDE 
\begin{eqnarray*}
-dY_{t}^{BS} &=&Z_{t}^{BS}\theta _{t}^{\top }dt-Z_{t}^{BS}dW_{t}, \\
Y_{T}^{BS} &=&\xi .
\end{eqnarray*}

\section{Pricing in incomplete markets}

\subsection{Upper and lower hedging prices}

In incomplete markets, there is no unique EMM and hence no unique method for
pricing a given contingent claim in an arbitrage-free way (in our setting, a
straightforward example of incomplete market is when $n<d$; see the works of
Cvitanic et al. \cite{cvitanic and kara},\cite{cvitanic and kara II},\cite%
{cvitanic and M} for more details).\ As is well known, the maximum arbitrage
free price of a contingent claim $\xi $ is given by%
\begin{equation*}
p_{t}^{up}:=\underset{\theta \in M}{\text{ess}\sup }E_{Q^{\theta }}[\xi
|\tciFourier _{t}],
\end{equation*}%
where $M$ is the set of all predictable $%
%TCIMACRO{\U{211d} }%
%BeginExpansion
\mathbb{R}
%EndExpansion
^{1\times d}$-valued processes $(\theta _{t})_{t\in \lbrack 0,T]}$ such that 
$E[\tint\nolimits_{0}^{T}\left\vert \theta _{t}\right\vert ^{2}dt]<+\infty $
and such that $S_{t}$ is a martingale with respect to $Q^{\theta }.$ $%
p_{t}^{up}$ is also referred to as the \textit{dynamic upper hedging price }%
of $\xi $ because it satisfies (Kunita \cite{Kunita}): 
\begin{equation*}
p_{t}^{up}=\text{ess}\inf \left\{ X_{x}^{(\pi )}(t)|\text{ }\exists \text{ }%
\pi \in \Pi \text{ s.t. }X_{x}^{(\pi )}(T)\geq \xi \text{, \ }P-a.s.\right\}
.
\end{equation*}

This means that $p_{t}^{up}$ represents the minimal wealth at time $t$
needed in order to be able to attain a terminal wealth $X^{(\pi )}(T)$\
which in no less than the guaranteed payoff $\xi $.

In order to fit into the setting of BSDE theory, we consider the set $%
M^{\prime }=\{\theta _{t}|\theta _{t}\in M$ such that $\left\vert \theta
_{t}\right\vert \leq u_{t},$ $t\in \lbrack 0,T]\},$ where $u_{t}$ is a
positive bounded deterministic process. Abusing notation, we will write $M$
for $M^{\prime }.$ The following theorem characterizes the dynamic upper
hedging price of $\xi $ as the solution of a BSDE:

\begin{theorem}[Chen and Wang \protect\cite{Chen and Wang}]
\label{chen and wang theo}Assume that $\xi \in L^{2}(\Omega ,\tciFourier
_{T},P)$. Then $p_{t}^{up}$ is the first component of the solution of BSDE 
\begin{eqnarray*}
-dY_{t} &=&u_{t}\left\vert Z_{t}\right\vert dt-Z_{t}dW_{t}, \\
Y_{T} &=&\xi .
\end{eqnarray*}
\end{theorem}

Similarly, the dynamic lower hedging price,

\begin{equation*}
p_{t}^{low}:=\underset{\theta \in M}{\text{ess}\inf }E_{Q^{\theta }}[\xi
|\tciFourier _{t}],
\end{equation*}%
satisfies%
\begin{equation*}
p_{t}^{low}=\text{ess}\sup \left\{ X_{x}^{(\pi )}(t)|\text{ }\exists \text{ }%
\pi \in \Pi \text{ s.t. }X_{x}^{(\pi )}(T)\leq \xi \text{, \ }P-a.s.\right\}
.
\end{equation*}%
It represents the maximal wealth allowed at time $t$ in order to attain a
terminal wealth $X^{(\pi )}(T)$\ which is no more than the promised payoff $%
\xi $. Using the same argument as Chen and Wang \cite{Chen and Wang}, one
can show the following theorem:

\begin{theorem}
\label{chen and wang low}Assume that $\xi \in L^{2}(\Omega ,\tciFourier
_{T},P)$. Then $p_{t}^{low}$ is the first component of the solution of BSDE 
\begin{eqnarray*}
-dY_{t} &=&-u_{t}\left\vert Z_{t}\right\vert dt-Z_{t}dW_{t}, \\
Y_{T} &=&\xi .
\end{eqnarray*}
\end{theorem}

If a trader charges the upper hedging price for selling an option, he can
trade to eliminate all risks.\ Likewise, if a trader buys an option for the
lower hedging price, he eliminates all risks.\ However, in general, the gap
between $p_{t}^{up}$\ and $p_{t}^{low}$\ is too wide to make either of them
a good candidate for the trading price in an incomplete market.

\subsection{Dynamic risk indifference pricing}

The starting point of the risk indifference pricing principle is a given
dynamic (or conditional) convex risk measure. Coherent risk measures were
introduced by Artzner et al.\ \cite{3} (see also Delbaen \cite{Delbaen I},%
\cite{Delbaen II}). Later, F\"{o}llmer and Schied \cite{13},\cite{Follmer
and Schied II}, and Frittelli and Rosazza Gianin \cite{15},\cite{Fritelli
and Gianin} introduced the class of convex risk measures.

The economic rationale behind the concept of risk measure is the following:
Let $\xi (\omega )\in L^{\infty }(\Omega ,\tciFourier _{T},P)$ be a
contingent bounded financial position at time $T$, then $\rho _{t,T}(\xi
)(\omega )$\ may be interpreted as the monetary degree of riskiness of $\xi $
when state $\omega $ occurs (for more details, see the above references, and
also F\"{o}llmer and Schied \cite{Folmmer schied book} or Detlefsen and
Scandolo \cite{Detlefsen}). Here follows the definition:

\begin{definition}[Convex risk measure]
\label{dynamic convex risk measures}A convex \textit{risk measure }$\rho
_{t,T}$ on $(\Omega ,\tciFourier _{T},P)$ \textit{conditional} to $(\Omega
,\tciFourier _{t},P)$ is a map $\rho _{t,T}:L^{\infty }(\Omega ,\tciFourier
_{T},P)\rightarrow L^{\infty }(\Omega ,\tciFourier _{t},P)$ satisfying the
following properties:

\begin{itemize}
\item[(a)] Monotonicity: $\forall \xi ,\eta \in L^{\infty }(\Omega
,\tciFourier _{T},P),$ if $\xi \leq \eta ,$ then $\rho _{t,T}(\xi )\geq \rho
_{t,T}(\eta ).$

\item[(b)] Translation invariance: $\forall \eta \in L^{\infty }(\Omega
,\tciFourier _{t},P),\forall \xi \in L^{\infty }(\Omega ,\tciFourier _{T},P)$%
, $\rho _{t,T}(\xi +\eta )=\rho _{t,T}(\xi )-\eta .$

\item[(c)] Convexity: $\forall \xi ,\eta \in L^{\infty }(\Omega ,\tciFourier
_{T},P),$ $\rho _{t,T}(\lambda \xi +(1-\lambda )\eta )\leq \lambda \rho
_{t,T}(\xi )+(1-\lambda )\rho _{t,T}(\eta ),$ for any $\lambda \in \lbrack
0,1].$

\item[(d)] Normalization: $\rho _{t,T}(0)=0.$

\item[(e)] Continuity from above: For any decreasing sequence $\left( \xi
_{n}\right) _{n\in 
%TCIMACRO{\U{2115} }%
%BeginExpansion
\mathbb{N}
%EndExpansion
}$ of elements of $L^{\infty }(\Omega ,\tciFourier _{T},P)$ such that $\xi
=\lim_{n}\xi _{n}$, the sequence $\rho _{t,T}(\xi _{n})$ has the limit $\rho
_{t,T}(\xi ).$

\item[(f)] Time consistency: for all $s,v$ such that $0\leq t\leq s\leq
v\leq T,$ $\rho _{t,v}(\xi )=\rho _{t,s}(\rho _{s,v}(\xi )),$ $\forall \xi
\in L^{\infty }(\Omega ,\tciFourier _{T},P).$

\item[(g)] $\forall \xi ,\eta \in L^{\infty }(\Omega ,\tciFourier _{T},P),$ $%
\forall A\in \tciFourier _{t},$ $\rho _{t,T}(\xi 1_{A}+\eta 1_{A^{c}})=\rho
_{t,T}(\xi )1_{A}+\rho _{t,T}(\eta )1_{A^{c}}.$

\item[(h)] $c_{t,T}(P):=$ess $\sup_{\xi \in L^{\infty }(\Omega ,\tciFourier
_{T},P)}\{E_{P}[-\xi |\tciFourier _{t}]-\rho _{t,T}(\xi )\}=0$ for any $t\in
\lbrack 0,T].$
\end{itemize}
\end{definition}

\bigskip

$\rho _{t,T}$ is called a \textit{normalized time-consistent dynamic convex
risk measure. }To simplify the notation, we often write $\rho _{t}$ instead
of $\rho _{t,T}$.

As in the utility indifference case (initiated by Hodges and Neuberger \cite%
{19}; see also Davis \cite{davis}, Davis et al. \cite{davis et al} and
Barles and Soner \cite{Barles and Soner}), the risk indifference pricing
principle starts from two situations:\bigskip

\textbf{(i)} If a person sells a contract which guarantees a payoff $\xi \in
L^{\infty }(\Omega ,\tciFourier _{T},P)$ at time $T$ and receives a payment $%
p_{t}$ for this, then at time $t$ the minimal risk involved for the seller is%
\begin{equation*}
\Phi _{t}^{\xi }(X_{x}^{(\pi )}(t)+p_{t})=\underset{\pi (\cdot )\in \Pi }{%
\text{ess}\inf }\rho _{t}(X_{x+p_{t}}^{(\pi )}(T)-\xi ).
\end{equation*}

\textbf{(ii)} If, on the other hand, no contract is sold, and hence no
payment is received, then at time $t$ the minimal risk for the person is%
\begin{equation*}
\Phi _{t}^{0}(X_{x}^{(\pi )}(t))=\underset{\pi (\cdot )\in \Pi }{\text{ess}%
\inf }\rho _{t}(X_{x}^{(\pi )}(T)).
\end{equation*}%
The dynamic risk indifference price is then defined as follows:

\begin{definition}[Seller's price]
The seller's dynamic risk indifference price $p_{t}=p_{t}^{seller}$ of a
claim $\xi \in L^{\infty }(\Omega ,\tciFourier _{T},P)$\ is the solution of
the equation%
\begin{equation}
\Phi _{t}^{\xi }(X_{x}^{(\pi )}(t)+p_{t})=\Phi _{t}^{0}(X_{x}^{(\pi )}(t)),
\label{E1}
\end{equation}%
for $t\in \lbrack 0,T].$ Thus $p_{t}^{seller}$ is the payment that makes a
person, at any time, risk indifferent between selling the contract with
liability $\xi $ and not selling the contract (and not receiving any payment
either).
\end{definition}

When $t=0$, $X_{x}^{(\pi )}(0)=x$ and (\ref{E1}) reduces to the static risk
indifference pricing problem of \O ksendal and Sulem \cite{25}.

Now, similarly, let 
\begin{equation*}
\Psi _{t}^{\xi }(X_{x}^{(\pi )}(t)+p_{t})=\underset{\pi (\cdot )\in \Pi }{%
\text{ess}\inf }\rho _{t}(\xi -X_{x+p_{t}}^{(\pi )}(T)),
\end{equation*}

and%
\begin{equation*}
\Psi _{t}^{0}(X_{x}^{(\pi )}(t))=\underset{\pi (\cdot )\in \Pi }{\text{ess}%
\inf }\rho _{t}(-X_{x}^{(\pi )}(T)).
\end{equation*}%
The buyer's dynamic risk indifference price is defined as follows:

\begin{definition}[Buyer's price]
The buyer's dynamic risk indifference price $p_{t}=p_{t}^{buyer}$ of a claim 
$\xi \in L^{\infty }(\Omega ,\tciFourier _{T},P)$\ is the solution of the
equation%
\begin{equation*}
\Psi _{t}^{\xi }(X_{x}^{(\pi )}(t)+p_{t})=\Psi _{t}^{0}(X_{x}^{(\pi )}(t)),
\end{equation*}%
for $t\in \lbrack 0,T].$ Thus $p_{t}^{buyer}$ is the payment that makes a
person, at any time, risk indifferent between buying the contract with
payoff $\xi $ and not buying the contract (and not making any payment
either).
\end{definition}

We first study in detail the case of the seller's risk indifference price.
By Bion-Nadal \cite{Bion nadal II} and Detlefsen and Scandolo \cite%
{Detlefsen}, it is known that under the assumptions above and in the setting
of a general filtration,

\begin{equation*}
\rho _{t,T}(\xi )=\underset{Q\in L}{\text{ess}\sup }\left\{ E_{Q}[-\xi
|\tciFourier _{t}]-c_{t,T}(Q)\right\} ,
\end{equation*}%
where%
\begin{equation*}
c_{t,T}(Q):=\underset{\xi \in L^{\infty }(\Omega ,\tciFourier _{T},P)}{\text{%
ess}\sup }\{E_{Q}[-\xi |\tciFourier _{t}]-\rho _{t,T}(\xi )\}
\end{equation*}%
is the penalty term associated to $\rho _{t,T}$, and%
\begin{equation*}
L=\{Q\text{ on }(\Omega ,\tciFourier _{T}):Q\sim P,Q=P\text{ on }\tciFourier
_{t}\}.
\end{equation*}

In particular, we have $c_{t}(Q):=c_{t,T}(Q)\geq 0.$ Taking this into
consideration, the problem of finding the risk indifference price $%
p_{t}^{risk}$ in (\ref{E1}) amounts to solving the following two zero-sum
stochastic differential game problems:

\begin{eqnarray}
\Phi _{t}^{\xi }(X_{x}^{(\pi )}(t)+p_{t}) &=&\text{ess}\underset{\pi (\cdot
)\in \Pi }{\inf }\underset{Q\in L}{\sup }\left\{ E_{Q}[-X_{x+p_{t}}^{(\pi
)}(T)+\xi |\tciFourier _{t}]-c_{t}(Q)\right\}  \notag \\
&&\text{and}  \label{E3} \\
\Phi _{t}^{0}(X_{x}^{(\pi )}(t)) &=&\text{ess}\underset{\pi (\cdot )\in \Pi }%
{\inf }\underset{Q\in L}{\sup }\left\{ E_{Q}[-X_{x}^{(\pi )}(T)|\tciFourier
_{t}]-c_{t}(Q)\right\} ,  \notag
\end{eqnarray}

for all $t\in \lbrack 0,T].$

\bigskip

These problems have a direct economic interpretation: Whilst the seller
tries to minimize the risk of the transaction over the set $\Pi $ of
admissible financial strategies, the market\ tries to maximize the corrected
expected loss over a set $L$ of \textquotedblleft generalized
scenarios\textquotedblright\ (i.e., probability measures $Q$), where
correction depends on scenarios.

We now use the following proposition of Delbaen et al. \cite{Delbaen et al}
in order to choose a representation of the penalty term $c_{t}$ such that
the pricing equation (\ref{E1}) holds for \textit{any} dynamic convex risk
measure.

\begin{proposition}[Delbaen, Peng and Rosazza Gianin \protect\cite{Delbaen
et al}]
\label{delb prop}Let $\rho _{t}$ be a dynamic convex risk measure satisfying
the assumption (a)-(h).\ Then, for any probability measure $Q$ equivalent to 
$P$, 
\begin{equation*}
c_{t}(Q)=E_{Q}\left[ \dint\nolimits_{t}^{T}f(u,\theta _{u})du|\tciFourier
_{t}\right]
\end{equation*}

for some suitable function $f:[0,T]\times \Omega \times 
%TCIMACRO{\U{211d} }%
%BeginExpansion
\mathbb{R}
%EndExpansion
^{d}\rightarrow \lbrack 0,+\infty ]$ such that $f(t,\omega ,\cdot )$ is
proper, convex, and lower-semicontinuous.
\end{proposition}

Let $N$ be the set of all predictable $%
%TCIMACRO{\U{211d} }%
%BeginExpansion
\mathbb{R}
%EndExpansion
^{1\times d}$-valued processes $(\theta _{t})_{t\in \lbrack 0,T]}$ such that 
$E[\tint\nolimits_{0}^{T}\left\vert \theta _{t}\right\vert ^{2}dt]<+\infty $
and such that $Q^{\theta }\in L.$ Based on proposition \ref{delb prop}, we
formulate the zero-sum stochastic differential game problem in (\ref{E3}) as
follows:

\begin{problem}
\label{saddle point pblm}Find $\Phi _{t}^{\xi }(X_{x}^{(\pi )}(t))$\ and an
optimal pair $(\widehat{\pi }(\cdot ),\widehat{\theta }(\cdot ))\in \Pi
\times N$\ such that%
\begin{equation*}
\Phi _{t}^{\xi }(X_{x}^{(\pi )}(t)):=\text{ess}\underset{\pi (\cdot )\in \Pi 
}{\inf }\underset{\theta (\cdot )\in N}{\sup }J_{t}(\pi ,\theta )=J_{t}(%
\widehat{\pi },\widehat{\theta }),
\end{equation*}%
and%
\begin{equation*}
J_{t}(\pi ,\theta ):=E_{Q_{\vartheta }}\left[ \xi -X_{x}^{(\pi
)}(T)-\dint\nolimits_{t}^{T}f(u,\theta _{u})du|\tciFourier _{t}\right] ,
\end{equation*}

for all $t\in \lbrack 0,T]$, and where $f$ is a predictable function
satisfying the assumptions of proposition \ref{delb prop}.
\end{problem}

\section{Upper and lower bounds on dynamic risk indifference prices}

In order to study general risk indifference prices, as well as upper and
lower hedging prices, in the same (BSDE) setting, we assume that the
following assumptions hold in the sequel:

\begin{itemize}
\item[(1)] $\xi \in L^{\infty }(\Omega ,\tciFourier _{T},P);$

\item[(2)] $\forall \theta _{t}\in N$, there exists a\ positive bounded
deterministic process $u_{t}$ such that $\left\vert \theta _{t}\right\vert
\leq u_{t},$ $\forall t\in \lbrack 0,T];$

\item[(3)] $\Pi $ is the set of strategies $\pi $\ uniformly bounded by a
constant $k.$

\item[(4)] $E[\tint\nolimits_{0}^{T}(f(t,\omega ,\theta
_{t}))^{2}dt]<+\infty $,

\item[(5)] $dt\times dP-a.s.,$ $\theta \rightarrow f(t,\omega ,\theta )$ is
continuously differentiable.
\end{itemize}

The following theorem gives sufficient conditions for solving problem \ref%
{saddle point pblm}:

\begin{theorem}
\label{sufficient cond}Let $\widehat{\pi }(\cdot )\in \Pi $ and $\widehat{%
\theta }(\cdot )\in N$ satisfy the following optimality conditions:%
\begin{equation}
X_{x}^{(\widehat{\pi })}(T)\geq X_{x}^{(\pi )}(T),\text{ }  \label{oc (a)}
\end{equation}

$\forall \pi (\cdot )\in \Pi ,$ $P-a.s.,$ and%
\begin{equation}
-f(t,\widehat{\theta }_{t})+z\widehat{\theta }_{t}^{\top }\geq -f(t,\theta
_{t})+z\theta _{t}^{\top },  \label{oc (b)II}
\end{equation}

$\forall z\in 
%TCIMACRO{\U{211d} }%
%BeginExpansion
\mathbb{R}
%EndExpansion
^{1\times d},\theta (\cdot )\in N,t\in \lbrack 0,T),$ $dt\times dP-a.s.$

Then, for any $t\in \lbrack 0,T],$ $J_{t}(\widehat{\pi },\widehat{\theta }%
)=Y_{t}^{\widehat{\theta },\widehat{\pi }},$ $P-a.s.,$ where $(Y_{t}^{%
\widehat{\theta },\widehat{\pi }},Z_{t}^{\widehat{\theta },\widehat{\pi }})$
is the solution of BSDE%
\begin{eqnarray}
-dY_{t}^{\widehat{\theta },\widehat{\pi }} &=&(-f(t,\widehat{\theta }%
_{t})+Z_{t}^{\widehat{\theta },\widehat{\pi }}\widehat{\theta }_{t}^{\top
})dt-Z_{t}^{\widehat{\theta },\widehat{\pi }}dW_{t},  \label{BSDE optimal} \\
Y_{T}^{\widehat{\theta },\widehat{\pi }} &=&\xi -X_{x}^{(\widehat{\pi })}(T),
\notag
\end{eqnarray}%
and $(\widehat{\pi }(\cdot ),\widehat{\theta }(\cdot ))$ is an optimal pair
for problem \ref{saddle point pblm}.
\end{theorem}

\begin{proof}
The existence of $\widehat{\pi }(\cdot )\in \Pi $ and $\widehat{\theta }%
(\cdot )\in N$ satisfying (\ref{oc (a)}) and (\ref{oc (b)II}) follows by
applying a predictable selection theorem (see, e.g., proposition 4.1 in Lim
and Quenez \cite{Lim and quenez}\ for a similar argument).\ Further, it is
direct to check that $J_{t}(\pi ,\theta )$ is equal to the unique solution $%
Y_{t}^{\theta ,\pi }$ of the following linear BSDE 
\begin{eqnarray}
-dY_{t}^{\theta ,\pi } &=&(-f(t,\theta _{t})+Z_{t}^{\theta ,\pi }\theta
_{t}^{\top })dt-Z_{t}^{\theta ,\pi }dW_{t},  \label{bsde theta} \\
Y_{T}^{\theta ,\pi } &=&\xi -X_{x}^{(\pi )}(T),  \notag
\end{eqnarray}

and that $J_{t}(\widehat{\pi },\widehat{\theta })$ is equal to the unique
solution $Y_{t}^{\widehat{\theta },\widehat{\pi }}$ of BSDE (\ref{BSDE
optimal}). Combining the optimality condition (\ref{oc (b)II}) with the
comparison theorem for BSDEs (El Karoui et al. \cite{9}), we have that $%
Y_{t}^{\theta ,\pi }\leq Y_{t}^{\widehat{\theta },\pi }$, $P-a.s.,$ from
which we deduce that%
\begin{equation*}
\text{ess}\underset{\pi (\cdot )\in \Pi }{\inf }\underset{\theta (\cdot )\in
N}{\sup }Y_{t}^{\theta ,\pi }\leq \text{ess}\underset{\theta (\cdot )\in N}{%
\sup }Y_{t}^{\theta ,\widehat{\pi }}\leq Y_{t}^{\widehat{\theta },\widehat{%
\pi }}.
\end{equation*}

Similarly, by (\ref{oc (a)}), we have $Y_{t}^{\widehat{\theta },\pi }\geq
Y_{t}^{\widehat{\theta },\widehat{\pi }}$, $P-a.s.,$ and 
\begin{equation*}
\text{ess}\underset{\pi (\cdot )\in \Pi }{\inf }\underset{\theta (\cdot )\in
N}{\sup }Y_{t}^{\theta ,\pi }\geq \text{ess}\underset{\pi (\cdot )\in \Pi }{%
\inf }Y_{t}^{\widehat{\theta },\pi }\geq Y_{t}^{\widehat{\theta },\widehat{%
\pi }}.
\end{equation*}

By uniqueness, it follows that%
\begin{equation*}
\text{ess}\underset{\pi (\cdot )\in \Pi }{\inf }\underset{\theta (\cdot )\in
N}{\sup }Y_{t}^{\theta ,\pi }=Y_{t}^{\widehat{\theta },\widehat{\pi }},\text{
}P-a.s.,
\end{equation*}

that is,%
\begin{equation*}
\text{ess}\underset{\pi (\cdot )\in \Pi }{\inf }\underset{\theta (\cdot )\in
N}{\sup }J_{t}(\pi ,\theta )=J_{t}(\widehat{\pi },\widehat{\theta }),\text{ }%
P-a.s.,
\end{equation*}

which concludes the proof.
\end{proof}

\bigskip

The following lemma will be very useful:

\begin{lemma}
\label{lemma max min}Assume that $\widehat{\theta }(\cdot )=\widehat{\theta }%
(\cdot ,\pi (\cdot ))\ $and $\widehat{\pi }(\cdot )$ satisfy the optimality
conditions (\ref{oc (a)}) and (\ref{oc (b)II}) $\forall \pi (\cdot )\in \Pi $%
, and assume that the function $\pi \rightarrow \theta (\pi )$ from $%
%TCIMACRO{\U{211d} }%
%BeginExpansion
\mathbb{R}
%EndExpansion
^{1\times n}$ into $%
%TCIMACRO{\U{211d} }%
%BeginExpansion
\mathbb{R}
%EndExpansion
^{1\times d}$ is a continuously differentiable function.\ Then $Q_{%
\widetilde{\theta }}$ is an EMM, where $\widetilde{\theta }(\cdot )=\widehat{%
\theta }(\cdot ,\widehat{\pi }(\cdot )).$
\end{lemma}

\begin{proof}
Consider the map 
\begin{eqnarray*}
J_{t}^{\pi ,\theta } &:&%
%TCIMACRO{\U{211d} }%
%BeginExpansion
\mathbb{R}
%EndExpansion
^{1\times d}\rightarrow L^{2}(\Omega ,\tciFourier _{t},dP\times dt) \\
&:&\theta \rightarrow E_{Q_{\vartheta }}\left[ \xi -X_{x}^{(\pi
)}(T)-\dint\nolimits_{t}^{T}f(u,\theta )du|\tciFourier _{t}\right] ,
\end{eqnarray*}%
and note that $J_{t}^{\pi ,\theta }=Y_{t}^{\theta }$ is the first component
of the solution of BSDE 
\begin{eqnarray*}
-dY_{t}^{\theta } &=&(-f(t,\theta )+Z_{t}^{\theta }\theta ^{\top
})dt-Z_{t}^{\theta }dW_{t}, \\
Y_{T}^{\theta } &=&\xi -X_{x}^{(\pi )}(T).
\end{eqnarray*}%
\ For a differentiable function $g:%
%TCIMACRO{\U{211d} }%
%BeginExpansion
\mathbb{R}
%EndExpansion
^{n}\rightarrow 
%TCIMACRO{\U{211d} }%
%BeginExpansion
\mathbb{R}
%EndExpansion
^{m}:x\rightarrow g(x),$ let $\nabla _{x}g$ denote the gradient matrix of $g$
with respect to $x$, i.e. $\left( \nabla _{x}g\right) _{i,j}=\frac{\partial
g^{j}(x)}{\partial x^{i}}$, for each $0\leq i\leq n,$ $0\leq j\leq m.$ By
proposition 2.4 of El Karoui et al. \cite{9} on the continuity and
differentiability of BSDEs with respect to a parameter (which may be
extended to multi-dimensional parameters)$,$ the map $\theta \rightarrow
Y_{t}^{\theta }$ is differentiable in $\theta $, with derivatives given by $%
\nabla _{\theta }Y_{t}^{\theta }$, the first component of the solution of
the following $d$-dimensional BSDE%
\begin{eqnarray*}
-d\nabla _{\theta }Y_{t}^{\theta } &=&(\nabla _{\theta }f(t,\theta
_{t})+(Z_{t}^{\theta })^{\top }+\nabla _{\theta }Z_{t}^{\theta }\theta
^{\top })dt-\nabla _{\theta }Z_{t}^{\theta }dW_{t}, \\
\nabla _{\theta }Y_{T}^{\theta } &=&\nabla _{\theta }(\xi -X_{x}^{(\pi )}(T).
\end{eqnarray*}%
The first order condition for a maximum point of the map $\theta \rightarrow
Y_{t}^{\theta }$ at $\widehat{\theta }(\cdot ,\pi (\cdot ))$ yields that $%
\nabla _{\theta }Y_{t}^{\theta }$ evaluated at $\widehat{\theta }=\widehat{%
\theta }(\cdot ,\pi (\cdot ))$ must be equal to zero, i.e. $\nabla _{\theta
}Y_{t}^{\widehat{\theta }}=0,$ almost surely, for all $t\in \lbrack 0,T],$
where $\nabla _{\theta }Y_{t}^{\widehat{\theta }}$ is the solution of%
\begin{eqnarray*}
-d\nabla _{\theta }Y_{t}^{\widehat{\theta }} &=&(\nabla _{\theta }f(t,%
\widehat{\theta }_{t})+(Z_{t}^{\widehat{\theta }})^{\top }+\nabla _{\theta
}Z_{t}^{\widehat{\theta }}\widehat{\theta }_{t}^{\top })dt-\nabla _{\theta
}Z_{t}^{\widehat{\theta }}dW_{t}, \\
\nabla _{\theta }Y_{T}^{\widehat{\theta }} &=&0.
\end{eqnarray*}

By Girsanov theorem, we can define the $Q_{\widehat{\theta }}$-Brownian
motion as $\widehat{W}_{t}:=W_{t}-\tint\nolimits_{0}^{T}\widehat{\theta }%
_{s}ds,$ and rewrite the above equation as 
\begin{eqnarray*}
-d\nabla _{\theta }Y_{t}^{\widehat{\theta }} &=&(\nabla _{\theta }f(t,%
\widehat{\theta }_{t})+(Z_{t}^{\widehat{\theta }})^{\top })dt-\nabla
_{\theta }Z_{t}^{\widehat{\theta }}d\widehat{W}_{t}, \\
\nabla _{\theta }Y_{T}^{\widehat{\theta }} &=&0.
\end{eqnarray*}

Taking the conditional expectation, we conclude that 
\begin{equation}
\nabla _{\theta }(f(t,\widehat{\theta }_{t})+(Z_{t}^{\widehat{\theta }%
})^{\top }=0,\text{ }dt\times dP-a.s.  \label{null theta}
\end{equation}%
Now, consider the map 
\begin{eqnarray*}
\widehat{J}_{t}^{\pi ,\widehat{\theta }} &:&%
%TCIMACRO{\U{211d} }%
%BeginExpansion
\mathbb{R}
%EndExpansion
^{1\times n}\rightarrow L^{2}(\Omega ,\tciFourier _{t},dP\times dt) \\
&:&\pi \rightarrow E_{Q_{\widehat{\vartheta }(\pi )}}\left[ \xi -X_{x}^{(\pi
)}(T)-\dint\nolimits_{t}^{T}f(u,\widehat{\theta }(\pi ))du|\tciFourier _{t}%
\right] ,
\end{eqnarray*}%
and note that $\widehat{J}_{t}^{\pi ,\widehat{\theta }}=Y_{t}^{\widehat{%
\theta }}.$ By the same argument, we must have that $\nabla _{\pi }Y_{t}^{%
\widehat{\theta }}$ evaluated at $\widetilde{\theta }=\widehat{\theta }%
(\cdot ,\widehat{\pi }(\cdot ))$ must be equal to 0, i.e.\ $\nabla _{\pi
}Y_{t}^{\widetilde{\theta }}=0,$ almost surely, for all $t\in \lbrack 0,T],$
where $\nabla _{\pi }Y_{t}^{\widetilde{\theta }}$ is the solution of the
following $n$-dimensional BSDE%
\begin{eqnarray*}
-d\nabla _{\pi }Y_{t}^{\widetilde{\theta }} &=&(\nabla _{\pi }\widetilde{%
\theta _{t}}(\nabla _{\theta }f(t,\widetilde{\theta }_{t})+(Z_{t}^{%
\widetilde{\theta }})^{\top })+\nabla _{\pi }Z_{t}^{\widetilde{\theta }}%
\widetilde{\theta }_{t}^{\top })dt-\nabla _{\pi }Z_{t}^{\widetilde{\theta }%
}dW_{t}, \\
\nabla _{\theta }Y_{T}^{\widetilde{\theta }} &=&\nabla _{\pi }(\xi
-X_{x}^{(\pi )}(T)).
\end{eqnarray*}

Again, by Girsanov theorem, we define the $Q_{\widetilde{\theta }}$-Brownian
motion as $\widetilde{W}_{t}:=W_{t}-\tint\nolimits_{0}^{T}\widetilde{\theta }%
_{s}ds$ and we obtain that 
\begin{eqnarray*}
-d\nabla _{\pi }Y_{t}^{\widetilde{\theta }} &=&\nabla _{\theta }(f(t,%
\widetilde{\theta }_{t})+Z_{t}^{\widetilde{\theta }}\widetilde{\theta }%
_{t}^{\top })\nabla _{\pi }\widetilde{\theta }dt-\nabla _{\pi }Z_{t}^{%
\widetilde{\theta }}d\widetilde{W}_{t}, \\
\nabla _{\theta }Y_{T}^{\widetilde{\theta }} &=&\nabla _{\pi }(\xi -%
\widetilde{X}_{x}^{(\pi )}(T)),
\end{eqnarray*}

where $\widetilde{X}$ is the portfolio process with respect to $\widetilde{W}%
.$ Combining the first order condition for a minimum point in $\widehat{\pi }
$ with (\ref{null theta}), we obtain that 
\begin{eqnarray*}
\nabla _{\pi }(\widetilde{X}_{x}^{(\pi )}(T)) &=&\nabla _{\pi
}(x+\dint\nolimits_{0}^{T}(\pi _{t}\mu _{t}S_{t}+\widetilde{\theta }%
_{t}\sigma _{t}\pi _{t}S_{t})dt+\dint\nolimits_{0}^{T}\sigma _{t}\pi
_{t}S_{t}d\widetilde{W}_{t}) \\
&=&0.
\end{eqnarray*}

Since the wealth process $X$ is almost surely nonnegative for any time $t\in
\lbrack 0,T],$ we conclude that $dt\times dP-a.s.,$ $\forall t\in \lbrack
0,T],$ 
\begin{equation*}
\mu _{t}+\widetilde{\theta }_{t}\sigma _{t}=0,
\end{equation*}

which concludes the proof.
\end{proof}

\bigskip

We can now show that the stochastic differential game problem \ref{saddle
point pblm}, which involves the supremum with respect to the set\ of all
probability measures $Q$ equivalent to $P$, can be reduced to a single
stochastic control problem with respect to the (narrower) set of EMM.

\begin{theorem}
\label{EMM theo}Assume that $\widehat{\theta }(\cdot )\ $and $\widehat{\pi }%
(\cdot )$ satisfy the assumptions of lemma \ref{lemma max min}. Then\ 
\begin{eqnarray*}
&&\text{ess}\underset{\pi (\cdot )\in \Pi }{\inf }\underset{\theta (\cdot
)\in N}{\sup }E_{Q_{\vartheta }}\left[ \xi -X_{x}^{(\pi
)}(T)-\dint\nolimits_{t}^{T}f(u,\theta _{u})du|\tciFourier _{t}\right] \\
&=&\text{ess}\underset{\theta (\cdot )\in M}{\sup }E_{Q_{\vartheta }}\left[
\xi -\dint\nolimits_{t}^{T}f(u,\theta _{u})du|\tciFourier _{t}\right] ,
\end{eqnarray*}%
for all $t\in \lbrack 0,T].$
\end{theorem}

\begin{proof}
By lemma \ref{lemma max min}, the probability measure $Q_{\widetilde{%
\vartheta }}$ induced by $\widetilde{\theta }(\cdot )=\widehat{\theta }%
(\cdot ,\widehat{\pi }(\cdot ))$ is an EMM, so we have 
\begin{eqnarray*}
&&\text{ess}\underset{\pi (\cdot )\in \Pi }{\inf }\underset{\theta (\cdot
)\in N}{\sup }E_{Q_{\vartheta }}\left[ \xi -X_{x}^{(\pi
)}(T)-\dint\nolimits_{t}^{T}f(u,\theta _{u})du|\tciFourier _{t}\right] \\
&=&\text{ess}\underset{\pi (\cdot )\in \Pi }{\inf }E_{Q_{\widehat{\vartheta }%
}}\left[ \xi -X_{x}^{(\pi )}(T)-\dint\nolimits_{t}^{T}f(u,\widehat{\theta }%
_{u})du|\tciFourier _{t}\right] \\
&=&E_{Q_{\widetilde{\vartheta }}}\left[ \xi -X_{x}^{(\widehat{\pi }%
)}(T)-\dint\nolimits_{t}^{T}f(u,\widetilde{\theta }_{u})du|\tciFourier _{t}%
\right] \\
&=&E_{Q_{\widetilde{\vartheta }}}\left[ \xi -\dint\nolimits_{t}^{T}f(u,%
\widetilde{\theta }_{u})du|\tciFourier _{t}\right] \\
&\leq &\text{ess}\underset{\theta (\cdot )\in M}{\sup }E_{Q_{\vartheta }}%
\left[ \xi -\dint\nolimits_{t}^{T}f(u,\theta _{u})du|\tciFourier _{t}\right]
.
\end{eqnarray*}

Conversely, since $M\subset N$, we always have that 
\begin{eqnarray*}
&&\text{ess}\underset{\pi (\cdot )\in \Pi }{\inf }\underset{\theta (\cdot
)\in N}{\sup }E_{Q_{\vartheta }}\left[ \xi -X_{x}^{(\pi
)}(T)-\dint\nolimits_{t}^{T}f(u,\theta _{u})du|\tciFourier _{t}\right] \\
&\geq &\text{ess}\underset{\pi (\cdot )\in \Pi }{\inf }\underset{\theta
(\cdot )\in M}{\sup }E_{Q_{\vartheta }}\left[ \xi -X_{x}^{(\pi
)}(T)-\dint\nolimits_{t}^{T}f(u,\theta _{u})du|\tciFourier _{t}\right] \\
&=&\text{ess}\underset{\theta (\cdot )\in M}{\sup }E_{Q_{\vartheta }}\left[
\xi -\dint\nolimits_{t}^{T}f(u,\theta _{u})du|\tciFourier _{t}\right] ,
\end{eqnarray*}

from which the claim follows.
\end{proof}

\bigskip

Define, $\forall \xi \in L^{\infty }(\Omega ,\tciFourier _{T},P),$%
\begin{equation*}
\rho _{t}^{M}(\xi ):=\text{ess}\underset{\theta (\cdot )\in M}{\sup }%
E_{Q_{\vartheta }}\left[ \xi -\dint\nolimits_{t}^{T}f(u,\theta
_{u})du|\tciFourier _{t}\right] .
\end{equation*}

The following corollary states that the buyer's and seller's risk
indifference prices can be formulated in terms of $\rho _{t}^{M}.$

\begin{corollary}
\label{corollary rho m}Let $\rho _{t}^{M}$ be defined as above.\ Then%
\begin{equation*}
p_{t}^{seller}=\rho _{t}^{M}(-\xi ),
\end{equation*}

and%
\begin{equation*}
p_{t}^{buyer}=-\rho _{t}^{M}(\xi ).
\end{equation*}
\end{corollary}

\begin{proof}
By definition, $p_{t}=p_{t}^{seller}$ is such that 
\begin{equation*}
\underset{\pi (\cdot )\in \Pi }{\text{ess}\inf }\rho _{t}(X_{x+p_{t}}^{(\pi
)}(T)-\xi )=\underset{\pi (\cdot )\in \Pi }{\text{ess}\inf }\rho
_{t}(X_{x}^{(\pi )}(T)).
\end{equation*}

By the translation invariance property of $\rho _{t}$, it follows that 
\begin{equation*}
p_{t}^{seller}=\underset{\pi (\cdot )\in \Pi }{\text{ess}\inf }\rho
_{t}(X_{x}^{(\pi )}(T)-\xi )-\underset{\pi (\cdot )\in \Pi }{\text{ess}\inf }%
\rho _{t}(X_{x}^{(\pi )}(T)).
\end{equation*}

The dual representation of dynamic convex risk measures and theorem \ref{EMM
theo} imply that%
\begin{equation*}
p_{t}^{seller}=\rho _{t}^{M}(-\xi )-\rho _{t}^{M}(0).
\end{equation*}

The first part of the claim follows because $\rho _{t}^{M}$ is a normalized
risk measure.\ Since the buyer's risk indifference price is defined as 
\begin{equation*}
\underset{\pi (\cdot )\in \Pi }{\text{ess}\inf }\rho _{t}(\xi
-X_{x+p_{t}}^{(\pi )}(T))=\underset{\pi (\cdot )\in \Pi }{\text{ess}\inf }%
\rho _{t}(-X_{x}^{(\pi )}(T)),
\end{equation*}%
a similar argument yields 
\begin{eqnarray*}
p_{t}^{seller} &=&-\left( \underset{\pi (\cdot )\in \Pi }{\text{ess}\inf }%
\rho _{t}(\xi -X_{x+p_{t}}^{(\pi )}(T))-\underset{\pi (\cdot )\in \Pi }{%
\text{ess}\inf }\rho _{t}(-X_{x}^{(\pi )}(T))\right) \\
&=&-(\rho _{t}^{M}(\xi )-\rho _{t}^{M}(0)) \\
&=&-\rho _{t}^{M}(\xi ).
\end{eqnarray*}
\end{proof}

\bigskip

Hence, in incomplete markets, risk indifference prices can be formulated in
terms of a risk measure $\rho _{t}^{M}$ which depends on the incompleteness
of the market $-$ since it involves the supremum with respect to the class
of EMM.\ 

\bigskip

Here is our main theorem:

\begin{theorem}
\label{main theorem}Let $p_{t}^{low}$ and $p_{t}^{up}$ be the lower and
upper hedging prices of $\xi $ defined in Section 3.1 and let $p_{t}^{buyer}$
and $p_{t}^{seller}$ be the buyer's and seller's risk indifference prices of 
$\xi $ defined is Section 3.2.\ Then, $P-a.s.$, $\forall t\in \lbrack 0,T],$%
\begin{equation*}
p_{t}^{low}\leq p_{t}^{buyer}\leq p_{t}^{seller}\leq p_{t}^{up}.
\end{equation*}
\end{theorem}

\begin{proof}
By theorem \ref{chen and wang theo}, we know that the upper heding price $%
p_{t}^{up}$ is equal to the solution $Y_{t}^{up}$\ of BSDE 
\begin{eqnarray*}
-dY_{t}^{up} &=&u_{t}\left\vert Z_{t}^{up}\right\vert dt-Z_{t}^{up}dW_{t}, \\
Y_{T}^{up} &=&\xi .
\end{eqnarray*}

On the other hand, by corollary \ref{corollary rho m}, 
\begin{eqnarray*}
p_{t}^{seller} &=&\rho _{t}^{M}(-\xi ) \\
&=&\text{ess}\underset{\theta (\cdot )\in M}{\sup }E_{Q_{\vartheta }}\left[
\xi -\dint\nolimits_{t}^{T}f(u,\theta _{u})du|\tciFourier _{t}\right] \\
&=&\text{ess}\underset{\theta (\cdot )\in M}{\sup }Y_{t}^{seller},
\end{eqnarray*}%
\ where $(Y_{t}^{seller},Z_{t}^{seller})$ is the solution of%
\begin{eqnarray*}
-dY_{t}^{seller} &=&(-f(t,\theta _{t})+Z_{t}^{seller}\theta _{t}^{\top
})dt-Z_{t}^{seller}dW_{t}, \\
Y_{T}^{seller} &=&\xi .
\end{eqnarray*}

By the comparison theorem for BSDEs (see El Karoui et al. \cite{9}), it
follows that $p_{t}^{seller}=\widehat{Y}_{t}^{seller}$, where 
\begin{eqnarray*}
-d\widehat{Y}_{t}^{seller} &=&(\text{ess}\underset{\theta (\cdot )\in M}{%
\sup }(-f(t,\theta _{t})+\widehat{Z}_{t}^{seller}\theta _{t}^{\top }))dt-%
\widehat{Z}_{t}^{seller}dW_{t}, \\
\widehat{Y}_{T}^{seller} &=&\xi .
\end{eqnarray*}

Since $f\in \lbrack 0,+\infty ]$ and $\left\vert \theta _{t}\right\vert \leq
u_{t},$ $-f(t,\theta _{t})+z\theta _{t}^{\top }\leq \left\vert z\right\vert
u_{t}$ $\forall z\in 
%TCIMACRO{\U{211d} }%
%BeginExpansion
\mathbb{R}
%EndExpansion
^{1\times d},$ and%
\begin{equation*}
\text{ess}\underset{\theta (\cdot )\in M}{\sup }(-f(t,\theta
_{t})+Z_{t}^{seller}\theta _{t}^{\top })\leq \left\vert
Z_{t}^{seller}\right\vert u_{t},\text{ }dt\times dP-a.s.
\end{equation*}

The comparison theorem for BSDEs implies that $Y_{t}^{up}\geq \widehat{Y}%
_{t}^{seller}$ $P-a.s.,$ which proves the last inequality.\ By theorem \ref%
{chen and wang low}, $p_{t}^{low}$ is equal to the solution $Y_{t}^{low}$\
of BSDE 
\begin{eqnarray*}
-dY_{t}^{low} &=&-u_{t}\left\vert Z_{t}^{low}\right\vert
dt-Z_{t}^{low}dW_{t}, \\
Y_{T}^{low} &=&\xi .
\end{eqnarray*}

On the other hand, by a similar argument, 
\begin{eqnarray*}
p_{t}^{buyer} &=&-\rho _{t}^{M}(\xi ) \\
&=&-\text{ess}\underset{\theta (\cdot )\in M}{\sup }E_{Q_{\vartheta }}\left[
-\xi -\dint\nolimits_{t}^{T}f(u,\theta _{u})du|\tciFourier _{t}\right] \\
&=&\text{ess}\underset{\theta (\cdot )\in M}{\inf }E_{Q_{\vartheta }}\left[
\xi +\dint\nolimits_{t}^{T}f(u,\theta _{u})du|\tciFourier _{t}\right] \\
&=&\text{ess}\underset{\theta (\cdot )\in M}{\inf }Y_{t}^{buyer},
\end{eqnarray*}%
where $(Y_{t}^{buyer},Z_{t}^{buyer})$ is the solution of 
\begin{eqnarray*}
-dY_{t}^{buyer} &=&(f(t,\theta _{t})+Z_{t}^{buyer}\theta _{t}^{\top
})dt-Z_{t}^{buyer}dW_{t}, \\
Y_{t}^{buyer} &=&\xi .
\end{eqnarray*}

By the comparison theorem for BSDEs, we have that $p_{t}^{buyer}=\widehat{Y}%
_{t}^{buyer}$, where 
\begin{eqnarray*}
-d\widehat{Y}_{t}^{buyer} &=&(\text{ess}\underset{\theta (\cdot )\in M}{\inf 
}(f(t,\theta _{t})+\widehat{Z}_{t}^{buyer}\theta _{t}^{\top }))dt-\widehat{Z}%
_{t}^{buyer}dW_{t}, \\
\widehat{Y}_{t}^{buyer} &=&\xi .
\end{eqnarray*}

Since $f(t,\theta _{t})+z\theta _{t}^{\top }\geq -\left\vert z\right\vert
u_{t}$ $\forall z\in 
%TCIMACRO{\U{211d} }%
%BeginExpansion
\mathbb{R}
%EndExpansion
^{1\times d},$ it follows from the comparison theorem for BSDEs that $%
Y_{t}^{low}\leq \widehat{Y}_{t}^{buyer}$ $P-a.s.$ It remains to prove the
second inequality, i.e.\ 
\begin{equation*}
\text{ess}\underset{\theta (\cdot )\in M}{\sup }Y_{t}^{seller}\geq \text{ess}%
\underset{\theta (\cdot )\in M}{\inf }Y_{t}^{buyer}.
\end{equation*}

This follows from the fact that%
\begin{eqnarray*}
&&\text{ess}\underset{\theta (\cdot )\in M}{\sup }Y_{t}^{seller}-\text{ess}%
\underset{\theta (\cdot )\in M}{\inf }Y_{t}^{buyer} \\
&\geq &\text{ess}\underset{\theta (\cdot )\in M}{\sup }\left(
Y_{t}^{seller}-Y_{t}^{buyer}\right) \\
&=&\text{ess}\underset{\theta (\cdot )\in M}{\sup }-2Y_{t}^{0} \\
&=&-2\text{ess}\underset{\theta (\cdot )\in M}{\inf }Y_{t}^{0}=0,
\end{eqnarray*}

where $(Y_{t}^{0},Z_{t}^{0})$ is the solution of 
\begin{eqnarray*}
-dY_{t}^{0} &=&f(t,\theta _{t})dt-Z_{t}^{0}d\widetilde{W}_{t}, \\
Y_{T}^{0} &=&0,
\end{eqnarray*}

and $\widetilde{W}$ is a $Q_{\theta }$-Brownian motion$.$
\end{proof}

\bigskip

Finally, we observe that in complete markets, i.e. when there is a unique
EMM, super-replication prices and risk indifference prices all reduce to the
single Black-Scholes price, that is, $%
p_{t}^{low}=p_{t}^{buyer}=p_{t}^{seller}=p_{t}^{up}=p_{t}^{BS},$ $\forall
t\in \lbrack 0,T].$

\end{document}